\documentclass[aps,preprint,nobibnotes,nofootinbib]{revtex4}

\usepackage{amsmath}

\begin{document}

\title{The Formulation of Quantum Field Theory in Curved Spacetime}

\author{Robert M. Wald \\ \it Enrico
Fermi Institute and Department of Physics \\ \it University of Chicago
\\ \it 5640 S.~Ellis Avenue, Chicago, IL~60637, USA}

\begin{abstract} The usual formulations of quantum field theory in Minkowski
spacetime make crucial use of Poincare symmetry, positivity of total
energy, and the existence of a unique, Poincare invariant vacuum
state. These and other key features of quantum field theory do not
generalize straightforwardly to curved spacetime. We discuss the
conceptual obstacles to formulating quantum field theory in curved
spacetime and how they can be overcome.
\end{abstract}

\maketitle

\bigskip
\bigskip

Quantum field theory in curved spacetime is a theory wherein matter is
treated fully in accord with the principles of quantum field theory,
but gravity is treated classically in accord with general
relativity. It is not expected to be an exact theory of nature, but it
should provide a good approximate description in circumstances where
the quantum effects of gravity itself do not play a dominant role.
Despite its classical treatment of gravity, quantum field theory in
curved spacetime has provided us with some of the deepest insights we
presently have into the nature of quantum gravity.

Quantum field theory as usually formulated contains many elements that
are very special to Minkowski spacetime. But we know from general
relativity that spacetime is not flat, and, indeed there are very
interesting quantum field theory phenomena that occur in
contexts---such as in the early universe and near black holes---where
spacetime cannot be approximated as nearly flat.

It is a relatively simple matter to generalize classical field theory
from flat to curved spacetime. That is because there is a clean
separation between the field equations and the solutions. The field
equations can be straightforwardly generalized to curved spacetime in
an entirely local and covariant manner. Solutions to the field
equations need not generalize from flat to curved spacetime, but this
doesn't matter for the formulation of the theory.

In quantum field theory, ``states'' are the analogs of ``solutions''
in classical field theory. However, properties of states---in
particular, the existence of a Poincare invariant vacuum state---are
deeply embedded in the usual formulations of quantum field theory in
Minkowski spacetime. For this reason and a number of other reasons, it
is highly nontrivial to generalize the formulation of quantum field
theory from flat to curved spacetime.

As a very simple, concrete example of a quantum field theory that
illustrates some key features of quantum field theory as well as some
of the issues that arise in generalizing the formlulation of quantum
field theory to curved spacetime, consider a free, real Klein-Gordon
field in flat spacetime,
\begin{equation}
\partial^a \partial_a \phi - m^2 \phi = 0 \,\, .
\end{equation}
The usual route towards formulating a quantum theory of $\phi$ is to
decompose it into a series of modes, and then treat each mode by the
rules of ordinary quantum mechanics.  To avoid technical awkwardness,
it is convenient to imagine that the field is in cubic box of side $L$
with periodic boundary conditions. We can then decompose $\phi$ as a
Fourier series in terms of the modes
\begin{equation}
\phi_{\vec{k}} \equiv L^{-3/2} \int 
e^{-i\vec{k} \cdot \vec{x}} \phi(t, \vec{x}) \,\, d^3 x \,\, 
\end{equation}
where $\vec{k} = \frac{2 \pi}{L} (n_1, n_2, n_3)$.
The Hamiltonian of the system is then given by
\begin{equation}
H = \sum_{\vec{k}} \frac{1}{2} \left(|\dot{\phi}_{\vec{k}}|^2 
+ \omega^2_{\vec{k}} |\phi_{\vec{k}}|^2 \right)
\end{equation}
where
\begin{equation}
\omega^2_{\vec{k}} = |\vec{k}|^2 + m^2 \,\, .
\end{equation}
Thus, a free Klein-Gordon field, $\phi$, in flat spacetime is
explicitly seen to be simply an infinite collection of decoupled
harmonic oscillators. If we take into account the fact that $\phi$ is
real but the modes $\phi_{\vec{k}}$ are complex, we find that each
$\phi_{\vec{k}}$ is given by the operator expression
\begin{equation}
\phi_{\vec{k}} = \frac{1}{2\omega_{\vec{k}}} (a_{\vec{k}} + a_{-\vec{k}}^\dagger) \,\, .
\end{equation}
where $a_{\vec{k}}$ and $a_{\vec{k}}^\dagger$ satisfy the usual
commutation relations
\begin{equation}
[a_{\vec{k}}, a_{\vec{k'}}] = 0 \,\, , \quad 
[a_{\vec{k}}, a_{\vec{k'}}^\dagger] = \delta_{\vec{k} \vec{k'}} I
\end{equation}
The Heisenberg field operator $\phi (t,\vec{x})$ is then formally given by
\begin{equation}
\phi (t, \vec{x}) = L^{-3/2} \sum_{\vec{k}} \frac{1}{2 \omega_{\vec{k}}}
\left(e^{i \vec{k} \cdot \vec{x} -i \omega_{\vec{k}} t} a_{\vec{k}} 
+ e^{-i \vec{k} \cdot \vec{x} + i \omega_{\vec{k}} t} a^\dagger_{\vec{k}}  
\right) \, .
\label{phih}
\end{equation}
However, this formula does {\it not} make sense as a definition of
$\phi$ as an operator at each point $(t, \vec{x})$. In essence, the
contributions from the modes at large $|\vec{k}|$ do
not diminish rapidly enough with $|\vec{k}|$ for the sum to
converge. However, these contributions are rapidly varying in
spacetime, so if we ``average'' the right side of eq.(\ref{phih}) in 
an appropriate manner over a
spacetime region, the sum will converge. More precisely,
eq.(\ref{phih}) defines $\phi$ as an ``operator valued distribution'',
i.e., for any (smooth, compactly supported) ``test function'', $f$,
the quantity
\begin{equation}
\phi(f) = \int f(t, \vec{x}) \phi (t, \vec{x}) d^4x 
\end{equation}
{\it is} well defined by eq.(\ref{phih}) if the integration is done
prior to the summation.

States of the free Klein-Gordon field are given the following
interpretation: The state, denoted $|0 \rangle$, in which all of the
oscillators comprising the Klein-Gordon field are in their ground
state is interpreted as representing the ``vacuum''. States of the
form $(a^\dagger)^n |0 \rangle$ are interpreted as ones where a total of $n$
``particles'' are present.

In an interacting theory, the state of the field may be such that the
field behaves like a free field at early and late times. In that case,
one has a particle interpretation at early and late times. The
relationship between the early and late time particle descriptions of
a state---given by the S-matrix---contains a great deal of the
dynamical information about the interacting theory.

The particle interpretation/description of quantum field theory in
flat spacetime has been remarkably successful---to the extent that one
might easily get the impression that, at a fundamental level, quantum
field theory is really a theory of ``particles''.  However, note that
even for a free field, the definition and interpretation of the
``vacuum'' and ``particles'' depends heavily on the ability to
decompose the field into its positive and negative frequency parts (as
can be seen explicitly from eq.(\ref{phih}) above). The ability to
define this decomposition makes crucial use of the presence of a time
translation symmetry in the background Minkowski spacetime. In a
generic curved spacetime without symmetries, there is no natural
notion of ``positive frequency solutions'' and, consequently, no
natural notion of a ``vacuum state'' or of ``particles''.

If one looks more deeply at the usual general formulations of quantum
field theory, it can be seen that many other properties that are
special to Minkowski spacetime are used in an essential way. This is
well illustrated by examining the Wightman axioms, since these axioms
abstract the key features of quantum field theory in Minkowski
spacetime in a mathematically clear way. We will focus attention on
the Wightman axioms below, but a similar discussion would apply to
other approaches, including the much less rigorous textbook treatments of
quantum field theory.

The Wightman axioms are the following \cite{sw}:
\begin{itemize}
\item The states of the theory are unit
rays in a Hilbert space, ${\mathcal H}$, that carries a unitary
representation of the Poincare group. 
\item 
The 4-momentum (defined by
the action of the Poincare group on the Hilbert space) is positive,
i.e., its spectrum is contained within the closed future light cone
(``spectrum condition'').
\item
There exists a unique, Poincare
invariant state (``the vacuum'').
\item
The quantum fields are
operator-valued distributions defined on a dense domain 
${\mathcal D} \subset {\mathcal H}$ that is both
Poincare invariant and invariant under the action of the fields and
their adjoints. 
\item
The fields transform in a covariant manner under
the action of Poincare transformations. 
\item
At spacelike separations,
quantum fields either commute or anticommute.
\end{itemize}

It is obvious that there are serious difficulties with extending the
Wightman axioms to curved spacetime, specifically:

\begin{itemize}

\item
A generic curved spacetime will not possess
any symmetries at all, so one certainly cannot require ``Poincare
invariance/covariance'' or invariance under any other type of
spacetime symmetry. 
\item
Even for a free quantum field, 
there exist unitarily inequivalent
Hilbert space constructions of the theory. For spacetimes with
a noncompact Cauchy surface---and in the absence of symmetries of the
spacetime---none appears ``preferred''.
\item
In a generic curved spacetime, there is no ``preferred'' choice of a
``vacuum state''.
\item
There is no analog of the spectrum condition in
curved spacetime that can be formulated in terms of the ``total
energy-momentum'' of the quantum field.

\end{itemize}
Thus, of all of the Wightman axioms, only the last one 
(commutativity or anticommutativity at spacelike separations) generalizes
straightforwardly to curved spacetime. 

I will now explain in more detail some of the difficulties associated
with generalizing the spectrum condition and the existence of a
preferred vacuum state to curved spacetime:

\medskip

\noindent
{\it Total energy in curved spacetime:} The stress energy tensor,
$T_{ab}$, of a classical field in curved spacetime is well defined and
it satisfies local energy-momentum conservation in the sense that
$\nabla^a T_{ab} = 0$. If $t^a$ is a vector field on spacetime
representing time translations and $\Sigma$ is a Cauchy surface, one can
define the total energy, $E$, of the field at ``time'' $\Sigma$ by
\begin{equation}
E = \int_\Sigma T_{ab} t^a n^b d\Sigma \,\, .
\label{E}
\end{equation}
Classically, for physically reasonable fields, the stress-energy
tensor satisfies the dominant energy condition, so $T_{ab} t^a n^b
\geq 0$. Thus, classically, we have $E
\geq 0$.  However, unless $t^a$ is a Killing field (i.e., unless the
spacetime is stationary), $E$ will not be conserved, i.e., independent
of choice of Cauchy surface, $\Sigma$.

In quantum field theory, it is expected that the stress-energy
operator will be well defined as an operator-valued distribution, and
it is expected to be conserved, $\nabla^a T_{ab} = 0$; see
\cite{hw4}. However, the definition of $T_{ab}$ requires spacetime
smearing. In Minkowski spacetime, since $E$ is conserved
one can, in effect, do ``time smearing'' without changing the value of
$E$, and there is a unique, well defined notion of total energy.
However, in the absence of time translation symmetry, one cannot
expect $E$ to be well defined at a ``sharp'' moment of time.  More
importantly, it is well known that $T_{ab}$ cannot satisfy the
dominant energy condition in quantum field theory (even if it holds
for the corresponding classical theory); locally, energy densities can
be arbitrarily negative. It is nevertheless true in Minkowski
spacetime that the total energy is positive for physically reasonable
states. However, in a curved spacetime without symmetries, there is no
reason to expect any ``time smeared'' version of $E$ to be
positive. Furthermore, there are simple examples with time translation
symmetry (such as a two-dimensional massless Klein-Gordon field in an
$S^1 \times {\bf R}$ universe) where $E$ can be computed explicitly
and is found to be negative \cite{bd}. Thus, it appears hopeless to
generalize the spectrum condition to curved spacetime in terms of the
positivity of a quantity representing ``total energy''.

\medskip

\noindent
{\it Nonexistence of a ``preferred vacuum state''and a notion of
``particles'':} As already noted above, for a free field in Minkowski
spacetime, the notion of ``particles'' and ``vacuum'' is intimately
tied to the notion of ``positive frequency solutions'', which, in turn
relies on the existence of a time translation symmetry. These notions
of a (unique) ``vacuum state'' and ``particles'' can be
straightforwardly generalized to (globally) stationary curved
spacetimes. However, there is no natural notion of ``positive
frequency solutions'' in a general, nonstationary curved spacetime.

Nevertheless for a free field on a general curved spacetime, a notion
of ``vacuum state'' can be defined as follows. A state is said to be
{\it quasi-free} if all of its $n$-point correlation functions
$\langle \phi(x_1) \dots \phi(x_n) \rangle$ can be expressed in terms
of its $2$-point correlation function by the same formula as holds for
the ordinary vacuum state in Minkowski spacetime.  A state is said to
be {\it Hadamard} if the singularity structure of its $2$-point
correlation function $\langle \phi(x_1) \phi(x_2) \rangle$ in the
coincidence limit $x_1 \rightarrow x_2$ is the natural generalization
to curved spacetime of the singularity structure of $\langle 0 |
\phi(x_1) \phi(x_2) |0 \rangle$ in Minkowski spacetime (see eq.(\ref{ope2})
below). Thus, in a general curved spacetime, the notion of a
quasi-free Hadamard state provides a notion of a ``vacuum state'',
associated to which is a corresponding notion of ``particles''. The
problem is that this notion of vacuum state is highly
non-unique. Indeed, for spacetimes with a non-compact Cauchy surface,
different choices of quasi-free Hadamard states give rise, in general,
to unitarily inequivalent Hilbert space constructions of the theory,
so in this case it is not even clear what the correct Hilbert space of
states should be. In the absence of symmetries or other special
properties of a spacetime, there does not appear to be any preferred
choice of quasi-free Hadamard state.

In my view, the quest for a ``preferred vacuum state'' in quantum
field theory in curved spacetime is much like the quest for a
``preferred coordinate system'' in classical general relativity. After
our more than 90 years of experience with classical general
relativity, there is a consensus that it is fruitless to seek a
preferred coordinate system for general spacetimes, and that the
theory is best formulated geometrically, wherein one does not have to
specify a choice of coordinate system in order to formulate the
theory. Similarly, after our more than 40 years of experience with
quantum field theory in curved spacetime, it seems similarly clear to
me that it is fruitless to seek a preferred vacuum state for general
spacetimes, and that the theory should be formulated in a manner that
does not require one to specify a choice specify a choice of state (or
representation) to define the theory.

\bigskip

Nevertheless, many of the above difficulties can be resolved in an
entirely satisfactory manner:

\begin{itemize}
\item
The difficulties that arise from the existence of unitarily
inequivalent Hilbert space constructions of quantum field theory in
curved spacetime can be overcome by formulating the theory via the
algebraic framework. The algebraic approach also fits in very well
with the viewpoint naturally arising in quantum field theory in curved
spacetime that the fundamental observables in quantum field theory are
the local quantum fields themselves.

\item
The difficulties that arise from the lack of an appropriate notion of the
total energy of the quantum field can be overcome by replacing the
spectrum condition by a ``microlocal spectrum condition'' that
restricts the singularity structure of the expectation values of the
correlation functions of the local quantum
fields in the coincidence limit.

\item
Many aspects of the requirement of Poincare invariance of the quantum
fields can be replaced by the requirement that the quantum fields be
locally and covariantly constructed out of the metric.
\end{itemize}

I will now explain these resolutions in more detail:

\medskip

\noindent
{\it The algebraic approach:} In the algebraic approach, 
instead of starting with a Hilbert space of
states and then defining the field observables as operators on this
Hilbert space, one starts with a *-algebra, ${\mathcal A}$, of field
observables.
A {\it state}, $\omega$, is simply a linear
map $\omega: {\mathcal A} \rightarrow {\bf C}$ that satisfies the
positivity condition $\omega (A^* A) \geq 0$ for all $A \in {\mathcal
A}$. The quantity $\omega(A)$ is interpreted as the expectation value
of the observable $A$ in the state $\omega$.

If $\mathcal H$ is a Hilbert space which carries a
representation, $\pi$, of $\mathcal A$, and if $\Psi \in {\mathcal H}$
then the map
$\omega: {\mathcal A} \rightarrow {\bf C}$ given by
\begin{equation}
\omega(A) = \langle \Psi | \pi (A) | \Psi \rangle 
\end{equation}
defines a state on $\mathcal A$ in the above sense.  Conversely, given
a state, $\omega$, on $\mathcal A$, we can use it to obtain a Hilbert
space representation of $\mathcal A$ by the following procedure, known
as the Gelfand-Naimark-Segal (GNS) construction. First, we use
$\omega$ to define a (pre-)inner-product on $\mathcal A$ by
\begin{equation}
(A_1, A_2) = \omega(A_1^* A_2) \, . 
\end{equation}
By factoring by zero-norm vectors and completing
this space, we get a Hilbert space $\mathcal H$, which carries
a natural representation, $\pi$, of $\mathcal A$. The vector $\Psi \in
{\mathcal H}$ corresponding to $I \in {\mathcal A}$ then satisfies
$\omega(A) = \langle \Psi | \pi (A) | \Psi \rangle$ for all $A \in
{\mathcal A}$.

Thus, the algebraic approach is not very different from the usual
Hilbert space approach in that every state in the algebraic sense
corresponds to a state in the Hilbert space sense and vice-versa. The
key difference is that, by adopting the algebraic approach, one may
simultaneously consider all states arising in all Hilbert space
constructions of the theory without having to make a particular choice
of representation at the outset. It is particularly important to
proceed in this manner in, e.g., studies of phenomena in the early
universe, so as not to prejudice in advance which states might be
present.

\medskip

\noindent
{\it The microlocal spectrum condition:} Microlocal analysis provides
a refined characterization of the singularities of a distribution by
examining the decay properties of its Fourier transform. More
precisely, let $D$ be a distribution on a manifold, $M$, and let
$(x,k_a)$ be a point in the cotangent bundle of $M$.  If $D$ has the
property that it can be multiplied by a smooth function, $f$, of
compact support with $f(x) \neq 0$, such that the Fourier transform of
$fD$ decays more rapidly than any inverse power of $|k|$ in a
neighborhood of the direction in Fourier transform space given by
$k_a$, then $D$ is said to be nonsingular at $(x,k_a)$. If $D$ does
not satisfy this property, then $(x,k_a)$ is said to lie in the {\it
  wavefront set} \cite{horm}, ${\rm WF}(D)$, of $D$. In the case of
quantum field theory in curved spacetime, the wavefront set can be
used to characterize the singular behavior of the distributions
$\omega [\phi_1 (x_1) \dots \phi_n (x_n)]$ (as a subset of the
cotangent bundle of $M \times \dots \times M$, where $M$ is the spacetime
manifold).

Now, for free fields in Minkowski spacetime, the positivity of total energy is
directly related to the choice of positive frequency solutions in the
decomposition (\ref{phih}). This, in turn, is directly
related to the ``locally positive frequency character'' of the
singular behavior of the $n$-point correlation functions $\omega [\phi
(x_1) \dots \phi (x_n)]$ in the coincidence limit. Consequently, it
can be shown that in Minkowski spacetime, the spectrum condition
(positivity of total energy) is equivalent to a {\it microlocal
spectrum condition} that restricts the wavefront set of $\omega
[\phi_1 (x_1) \dots \phi_n (x_n)]$. This microlocal spectrum condition
can be generalized straightforwardly to curved spacetime. In this
manner, it is possible to impose the requirement that states have a
``locally positive frequency character'' even in spacetimes where
there is no natural global notion of ``positive frequency'' (i.e., no
global notion of Fourier transform).

\medskip

\noindent
{\it Local and covariant fields}: It is often said that in special
relativity one has invariance under ``special coordinate
transformations'' (i.e., Poincare transformations), whereas in general
relativity, one has invariance under ``general coordinate
transformations'' (i.e., all diffeomorphisms). However, this is quite
misleading. By explicitly incorporating the flat spacetime metric into
the formulation of special relativity, it can easily be seen that
special relativity can be formulated in as ``generally covariant'' a
manner as general relativity, but the act of formulating special
relativity in a generally covariant manner does not provide one with
any additional symmetries or other useful conditions. The true meaning
of ``general covariance'' is that the theory is constructed in a local
manner from the spacetime metric and other dynamical fields, with no
non-dynamical background structure (apart from manifold structure, and
choices of space and time orientations and spin structure) playing any
role in the formulation of the theory. This is the proper
generalization of the notion of Poincare invariance to general
relativity.

In the present context, we wish to impose the requirement that in an
arbitrarily small neighborhood of a point $x$, the quantum fields
$\Phi$ under consideration ``be locally and covariantly constructed
out of the spacetime geometry'' in that neighborhood. In order to
formulate this requirement, it is essential that quantum field theory
in curved spacetime be formulated for {\it all} (globally hyperbolic)
curved spacetimes---so that we can formulate the notion that ``nothing
happens'' to the fields near $x$ when we vary the metric in an
arbitrary manner away from point $x$.  The notion of a local and
covariant field may then be formulated as follows \cite{bfv}: Suppose that we
have a causality preserving isometric embedding $i: M \rightarrow
{\mathcal O'} \subset M'$ of a spacetime $(M, g_{ab})$ into a region
$\mathcal O'$, of a spacetime $(M', g'_{ab})$.  We require that this
embedding induce a natural isomorphism of the quantum field algebra
${\mathcal A} (M)$ of the spacetime $(M, g_{ab})$ and the subalgebra
of the quantum field algebra ${\mathcal A} (M')$ associated with
region $\mathcal O'$. We further demand that under this isomorphism,
each quantum field $\Phi(f)$ on $M$ be taken into the corresponding
quantum field $\Phi(i^* f)$ in $\mathcal O'$.

In what sense is this condition a replacement for Poincare covariance?
In the case of Minkowski spacetime, we can isometrically embed all of
Minkowski spacetime into itself by a Poincare transformation. The
above condition then provides us with an action of the Poincare group on
the field algebra of Minkowski spacetime and also requires each
quantum field in Minkowski spacetime to transform covariantly under
Poincare transformations. Thus, the above condition contains much of the
essential content of Poincare invariance, but it is applicable to
arbitrary curved spacetimes without symmetries.

\bigskip

Let us now take stock of where things stand on the generalization of
the basic principles of quantum field theory---as expressed by the
Wightman axioms---to curved spacetime.

\begin{itemize}
\item The axiom that requires states to lie in a Hilbert space that
carries a unitary representation of the Poincare group is
satisfactorily replaced by formulating theory via the algebraic approach
and requiring that the quantum fields be local and covariant.
\item 
The spectrum condition is satisfactorily replaced by the microlocal
spectrum condition.
\item
The axiom stating that quantum fields are operator-valued
distributions defined on a dense domain that is Poincare invariant and
invariant under the action of the fields and their adjoints is
satisfactorily replaced by the GNS construction in the algebraic approach
and the local and covariant field condition.
\item
The axiom that the fields transform in a covariant manner under the action of
Poincare transformations is satisfactorily replaced by the local and
covariant field condition.
\item
As previously noted, the condition that at spacelike separations
quantum fields either commute or anticommute generalizes
straightforwardly to curved spacetime.
\end{itemize}

Thus, the only Wightman axiom that does not admit a satisfactory
generalization to curved spacetime based on the ideas described above is the
existence of a unique, Poincare invariant state (``the vacuum'').
This axiom plays a key role in the proofs of the PCT and
spin-statistics theorem and many other results, so one would lose a
great deal of the content of quantum field theory if one simply
omitted this axiom. In particular, vacuum expectation values of
products of fields play an important role in many arguments, and it is
crucial that these ``c-number'' quantities share the symmetries of the
fields. On the other hand, we have already argued that it is hopeless
to define a unique ``preferred state'' in generic spacetimes.
Furthermore, states are inherently global in character and cannot
share the ``local and covariant'' property of fields.

What is the appropriate replacement in curved spacetime of the
requirement that there exist a Poincare invariant state in Minkowski
spacetime? Hollands and I have recently proposed \cite{hw5} that the
appropriate replacement is the existence of an operator product
expansion of the quantum fields.  An {\it operator product expansion}
(OPE) is a short-distance asymptotic formula for products of quantum
fields near point $y$ in terms of quantum fields defined at $y$, 
i.e., formulae of the form
\begin{equation}
\phi^{(i_1)}(x_1) \cdots \phi^{(i_n)}(x_n) \sim \sum_{(j)} C^{(i_1)
\dots (i_n)}_{(j)}(x_1, \dots, x_n; y) \, \phi^{(j)}(y)
\label{ope}
\end{equation}
for all $i_1, \dots, i_n$, which hold as asymptotic
relations as $\{x_1, \dots, x_n \} \rightarrow y$.  
The
simplest example of an OPE is that for a product of two free Klein-Gordon
fields in curved spacetime. We have
\begin{equation}
\phi(x_1) \phi(x_2) = H(x_1,x_2) {\bf 1} + \phi^2 (y) + \dots
\label{ope2}
\end{equation}
where $H$ is a locally and covariantly constructed Hadamard
distribution (see, e.g., \cite{kw} or \cite{wald}
for the precise form of $H$) and
``$\dots$'' has higher scaling degree than the other terms (i.e., it
goes to zero more rapidly in the limit $x_1,x_2 \rightarrow y$). An
OPE exists for free fields in curved spacetime and Hollands \cite{h}
has shown that it holds order-by-order in perturbation theory for
renormalizable interacting fields in curved spacetime. The requirement
that an operator product expansion exists and satisfies a list of
suitable properties \cite{hw5} appears to provide an appropriate
replacement for the requirement of the existence of a Poincare invariant
state. In particular, the distributional coefficients of the identity
element in OPE expansions play much of the role played by ``vacuum
expectation values'' in Minkowski spacetime quantum field theory.

By elevating the existence of an OPE to a fundamental status, Hollands
and I have been led to the following viewpoint on quantum field theory
in curved spacetime: The background structure, $\mathcal M$, of
quantum field theory in curved spacetime is the spacetime $(M,
g_{ab})$, together with choices of time orientation, spacetime
orientation, and spin structure. For each $\mathcal M$, we have an
algebra ${\mathcal A}({\mathcal M})$ of local field observables. In
traditional algebraic approaches to quantum field theory, ${\mathcal
  A}({\mathcal M})$ would contain the full information about the
quantum field theory. However, in our approach, ${\mathcal A}({\mathcal M})$, is
essentially nothing more than the free *-algebra generated by the list
of quantum fields $\phi^{(i)}(f)$ (including ``composite fields''),
though it may be factored by relations that arise from the OPE.

In our viewpoint, all of the nontrivial information about the quantum
field theory is contained in its OPE, eq.(\ref{ope}). The distributions
$C^{(i_1) \dots (i_n)}_{(j)}(x_1, \dots, x_n; y)$ appearing in
eq.(\ref{ope}) are required to satisfy a list of properties, which
include locality and covariance and an ``associativity''
property. States are positive linear maps on the algebra that satisfy
the OPE relations as well as microlocal spectrum conditions.

Spin-statistics and PCT theorems have been proven within this
framework \cite{hw5}.  Interestingly, the PCT theorem relates processes in a
given spacetime to processes (involving charge conjugate fields) in a
spacetime with the opposite time orientation, e.g., it relates
processes involving particles in an expanding universe to processes
involving antiparticles in a contracting universe.

Renormalized perturbation theory can be carried out within this
framework \cite{bf}, \cite{hw1}, \cite{hw2}. 
For a free field, composite fields (Wick powers)---such as
$\phi^2$ and $T_{ab}$---and all time-ordered products of fields can be
defined in local and covariant manner. (However, ``normal ordering'' 
cannot be used to define composite fields.) The definition
of time-ordered-products 
is unique up to ``renormalization ambiguities'' of the
type expected from Minkowski spacetime analyses, but with additional
local curvature ambiguities (which also occur in the definition of the
composite fields). Theories that are renormalizable in
Minkowski spacetime remain renormalizable in curved spacetime.
Renormalization group flow can be defined in
terms of the behavior of the quantum field theory under scaling of the
spacetime metric, $g_{ab} \rightarrow \lambda^2 g_{ab}$ \cite{hw3}. 
Additional renormalization conditions can
be imposed so that, in perturbation theory,
the stress-energy tensor of the interacting field is
conserved (for an arbitrary covariant interaction) \cite{hw4}.

In summary, the attempt to describe quantum field phenomena in curved
spacetime has directly led to a viewpoint where symmetries and notions
of ``vacuum'' and ``particles'' play no fundamental role. The theory
is formulated in a local and covariant manner in terms of the quantum
fields.  This formulation is very well suited to investigation of
quantum field effects in the early universe. In addition, the
definition of composite fields, such as the stress-energy tensor, is
intimately related to the OPE, and thus arises naturally in this
framework.  It is my hope that quantum field theory in curved
spacetime will continue to provide us with deep insights into the
nature of quantum phenomena in strong gravitational fields {\it and}
into the nature of quantum field theory itself.

\bigskip

\noindent {\bf Acknowledgments}

This research was supported in part by NSF grant PHY04-56619 to the
University of Chicago.

\end{document}